# Electrolytic Conversion of Bicarbonate into CO in a Flow Cell

Tengfei Li,[1‡] Eric W. Lees,[2‡] Maxwell Goldman,[1]

Danielle A. Salvatore,[2] David M. Weekes,[3] Curtis P. Berlinguette[1,2,3*]

[1]Department of Chemistry, The University of British Columbia, 2036 Main Mall, Vancouver, BC V6T 1Z1, Canada

[2]Department of Chemical & Biological Engineering, The University of British Columbia, 2360 East Mall, Vancouver, BC V6T 1Z3, Canada

[3]Stewart Blusson Quantum Matter Institute, The University of British Columbia, 2355 East Mall, Vancouver, BC V6T 1Z4, Canada

Correspondence: cberling@chem.ubc.ca

[‡] T.L. and E.W.L. contributed equally

**Summary**

Electrochemical $CO_2$ reduction offers a method to use renewable electricity to convert $CO_2$ into CO and other carbon-based chemical building blocks. While nearly all studies rely on a $CO_2$ feed, we show herein that aqueous $HCO_3^-$ solutions can also be electrochemically converted into CO gas at meaningful rates in a flow cell. We achieved this result in a flow cell containing a bipolar membrane (BPM) and a silver nanoparticle catalyst on a porous carbon support. Electrolysis upon a $N_2$-saturated 3.0-M $KHCO_3$ electrolyte solution yields CO with a faradaic efficiency ($F.E._{CO}$) of 81% at 25 mA cm$^{-2}$ and 37% at 100 mA cm$^{-2}$. This output is comparable to the analogous experiment where the electrolyte is saturated with gaseous $CO_2$ (faradaic efficiency for CO is 78% at 25 mA cm$^{-2}$ and 35% at 100 mA cm$^{-2}$). The $H^+$ flux from the BPM is critical to this chemistry in that it reacts with the $HCO_3^-$ feed to generate $CO_2$, which is then reduced to CO at the gas diffusion electrode. These results are important in that they show that the addition of gaseous $CO_2$ to $HCO_3^-$ electrolytes is not necessary in order to obtain reduced carbon products with a flow cell architecture. This process offers a means of using electrolysis to bypass the thermally-intensive step of extracting $CO_2$ from $HCO_3^-$ solutions generated in carbon capture schemes.

**Keywords**

## Introduction

A number of different cell configurations have been proposed in an effort to reach the high current densities and product selectivities that would be required by a commercial $CO_2$ electrolyzer, yet there is no low-temperature $CO_2$ electrolyzer available to the market.[1–7] The majority of cell configurations supply $CO_2$ to the cathode of the electrochemical cell as a saturated solution in a supporting aqueous electrolyte.[1,2,7] There are also many examples where gaseous $CO_2$ is delivered directly to the electrode without a liquid medium.[8–12] A challenge with $CO_2$-fed liquid systems is that bubbling gaseous $CO_2$ into the electrolyte reaches a saturated $[CO_2]$ of ~0.033 M at ambient temperature and pressure prior to and during electrolysis. This saturation point fundamentally limits the maximum current density that can be achieved for $CO_2$ reduction in the bulk liquid phase.[13] Furthermore, the addition of $CO_2$ to the electrolyte causes a lowering in pH resulting in conditions that favour the competing hydrogen evolution reaction (HER) instead of the $CO_2$ reduction reaction ($CO_2$RR).[14–16] These characteristics of $CO_2$-saturated electrolyte systems provide the impetus to explore alternative pathways for accessing electrolytically-reduced carbon products at high current densities.

A system capable of efficiently utilizing bicarbonate ($HCO_3^-$) as a source of carbon to form electrochemical products is compelling because the maximum carbon concentration in a saturated aqueous solution of $KHCO_3$ (~3.3 M) is greater than for saturated $CO_2$ (33 mM). The ability to convert $HCO_3^-$ in solution to a reduced carbon product would also provide an opportunity to avoid electrolyte acidification caused by the addition of gaseous $CO_2$. However, very few examples of systems capable of these electrochemical transformations currently exist. A study by Hori in 1983 showed that formate could be produced from the electrochemical reduction of a 1.0-M $NaHCO_3$ solution without a $CO_2$ supply in an H-cell architecture containing a mercury catalyst, but with a partial current density for formate ($J_{formate}$) of <1 mA cm$^{-2}$.[17] Kanan and coworkers reported in 2015 that a palladium catalyst in an H-cell was able to reduce 2.8-M $KHCO_3$ into formate without a $CO_2$ feed at $J_{formate}$ = 3.2 mA cm$^{-2}$ (*c.f.* 6.1 mA cm$^{-2}$ in a $CO_2$-fed $HCO_3^-$ solution). To our knowledge, these results set the benchmark for electrochemical $HCO_3^-$ conversion to a reduced carbon product in a system without supply of $CO_2$.[14] All other reports of $HCO_3^-$ reduction chemistry show lower current densities and faradaic efficiencies compared to $CO_2$-fed systems. Moreover, formate is the only product that has been reported from the reduction of $HCO_3^-$ without a $CO_2$ feed, thus the electrochemical reduction of $HCO_3^-$ into CO without a $CO_2$ supply is not currently known.

Our program has an interest in developing flow cells that mediate $CO_2RR$ at high current densities.[10] Our previous studies have shown that a membrane reactor architecture with a BPM separating the cathode and the anode gas diffusion layers is effective at mediating $CO_2RR$ with $CO_2$ delivered either as a gas[8] or as dissolved in aqueous electrolyte.[9] While a gaseous $CO_2$ feed enables higher current densities,[10,18,19] there are practical advantages to delivering liquid rather than gas to the reactor.[10] However, realizing high current densities with a liquid feed is challenging because of the low solubility limits of $CO_2$. Recognizing that the BPM delivers $H^+$ to the cathode, we became intrigued by the possibility that the high $H^+$ flux at the BPM could react with $HCO_3^-$ to form a high local concentration of $CO_2$ that would be available for electrochemical reduction at the catalyst,[20,21] and without the need for an external supply of $CO_2$.

Following this line of inquiry, we built an electrochemical flow reactor[8,9] to electrochemically reduce $KHCO_3$ solutions without the need to supply the electrolyte with gaseous $CO_2$. We demonstrate herein a 3.0-M $KHCO_3$ electrolyte yeilds a F.E.$_{CO}$ of 81% at 25 mA cm$^{-2}$ that is comparable with a system wherein the electrolyte is saturated with $CO_2$ gas (F.E.$_{CO}$ = 78%). This experiment provides the following unique observations: (i) a carbon product other than formate as the product of electrochemical reduction performed upon a $HCO_3^-$ solution; (ii) comparable electrolytic activity measured for $HCO_3^-$ electrolytes with and without saturation from a gaseous $CO_2$ feed; (iii) the electrochemical conversion of $HCO_3^-$ solutions in a continuous flow reactor; and (iv) that the BPM in flow cell enhances the conversion of $HCO_3^-$ to CO through a $CO_2$ mediated pathway. These results illuminate an alternative way to electrolytically form carbon-based products derived from $CO_2$. This methodology also presents a new opportunity for air capture schemes because electrolysis could potentially bypass the energy-intensive thermal extraction of $CO_2$ gas from $HCO_3^-$ solutions.[22]

## Results and discussion

A two-electrode liquid flow-cell electrochemical reactor was used for our electrochemistry experiments (Figure 1).[8] The flow cell was built in-house and consists of housing, gaskets, anode and cathode flow-field plates and a membrane electrode assembly (MEA). The anode and cathode housings are made from stainless steel and serve to deliver liquid electrolyte to the anode and cathode. The titanium cathode flow plate and stainless steel anode flow plate sandwich the 4-cm$^2$ MEA. The anode (316 stainless steel) and cathode (grade 2 titanium) flow-field plates (active areas = 4 cm$^2$) contain serpentine channels 1.5 mm wide and 1.5 mm deep with 1-mm ribs. The MEA consists of a nickel foam anode ($2.5 \times 2.5$ cm$^2$), a BPM ($3 \times 3$ cm$^2$), and a silver nanoparticle-coated porous carbon support as the cathode (2.5 x 2.5 cm$^2$). The entire assembly is sandwiched between the two stainless housings fastened with 8 bolts. The silver-coated carbon cloth cathode was prepared by spray coating carbon cloth (CeTech) with a mixture of commercially-available silver nanopowder and Nafion$^{TM}$ solution. The anode feed was 1.0-M KOH delivered at 50 mL min$^{-1}$ and the cathode feed was an aqueous solution of either 3.0-M K$_2$CO$_3$ or 3.0-M KHCO$_3$ delivered at 50 mL min$^{-1}$ continuously purged through the head space with either N$_2$ or sparged with CO$_2$ gas at a rate of 160 sccm. Cyclic voltammetry measurements of the two-electrode flow cell were collected over the -1.5 to -3.5 V range (reported as the cell potential). Faradaic efficiencies for CO production were measured at constant cathodic current densities of 25, 50, 75 and 100 mA cm$^{-2}$. pH measurements were obtained by diverting ~10 mL of catholyte flowing through the peristaltic pump into a glass vial from which the pH was measured. Gas chromatography (GC) analysis of the gaseous headspace at the electrolyzer outlet showed CO, CO$_2$, and H$_2$ as the only detectable products. The cathode solution was analyzed by [1]H NMR after electrolysis and no liquid products were detected. F.E. for CO is reported with the remainder of the electrochemical products confirmed as H$_2$. Further description of the setup and full cell dimensions are included in the Supplementary Information (Figure S1). Control experiments were also applied using an analogous flow cell architecture with an anion exchange membrane (AEM) in place of the BPM.

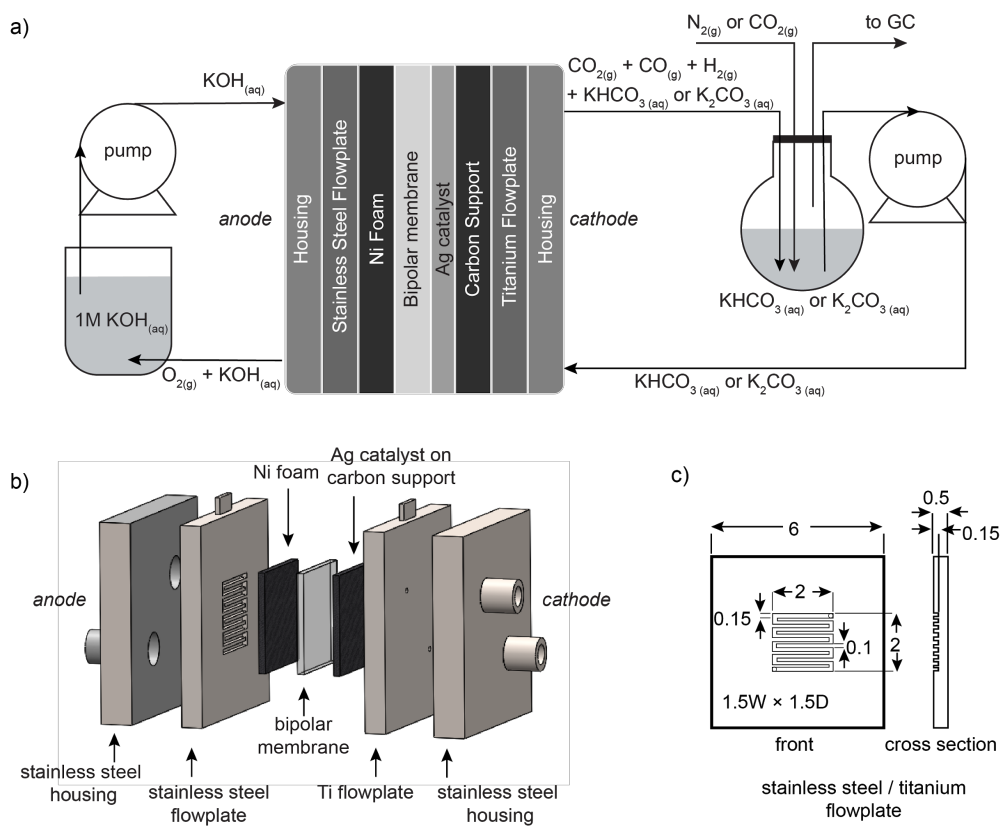

**Figure 1.** Experimental setup of (a) electrochemical flow cell experiment (b) an expanded view of the flow cell, and (c) dimensions of the cathode & anode flowplates. Nickel foam (anode) and silver deposited on a porous carbon support (cathode) were separated by a BPM. 1.0-M KOH electrolyte was circulated through the stainless flow plate and oxidized into $O_2$ gas at the anode. Either $KHCO_3$ or $K_2CO_3$ electrolyte solutions with $CO_2$ sparged in or with $N_2$ purged through the headspace were circulated through the titanium flow plate and reduced into CO at the cathode. The cathodic products were analyzed by gas chromatography (GC). Electrolyte flows were driven by peristaltic pumps at 50 mL min$^{-1}$. Gas flows ($N_2$ or $CO_2$) were set at 160 sccm. All dimension in (d) are in cm.

### *Electrolysis of $HCO_3^-$ solutions produce CO*

The flow cell setup described above was used to test the electrochemical production of CO during electrolysis of two 3.0-M $KHCO_3$ solution systems: (i) $KHCO_3$ reservoir sparged with $CO_2$; and (ii) $N_2$ purged through the headspaces of the KHCO$_3$ reservoirs. Cyclic voltammograms (CVs) were collected between potentials of -1.0 and -3.5 V (Figure 2a) in a two-electrode flow cell, and F.E.$_{CO}$ values were measured between current densities of 25 and 100 mA cm$^{-2}$ in 25-mA cm$^{-2}$ increments (Figure 2b). The viability of the flow cell towards CO$_2$ reduction was confirmed by results from the CO$_2$-saturated 3.0-M KHCO$_3$ solution (Figure 2, black): The CV exhibits a sharp rise in current density at -2.5 V and a current density of 90 mA cm$^{-2}$ at -3.5 V was measured. The moderate F.E.$_{CO}$ of 78% exhibited at low current densities (25 mA cm$^{-2}$) is further reduced at higher current densities (e.g., 35% at 100 mA cm$^{-2}$). These results are consistent with previous reports of CO$_2$-to-CO reduction in liquid-fed flow cells.[9,23,24]

Having validated the CO$_2$RR activity of our flow cell, we then investigated the electrochemical reduction of HCO$_3^-$ in the absence of a CO$_2$ supply. CVs collected over the -1.0 to -3.5 V cell potential range show similar reductive sweep profiles to the CO$_2$-saturated solution (Figure 2a). Peak current density of 100 mA cm$^{-2}$ was measured for the KHCO$_3$ solution in the absence of CO$_2$ supply. Electrochemical reduction of N$_2$-saturated 3.0-M KHCO$_3$ solution showed a F.E.$_{CO}$ of 81% at a current density of 25 mA cm$^{-2}$, and 37% at 100 mA cm$^{-2}$ (Figure 2b, orange). These results represent the first observations of the production of CO from a HCO$_3^-$ solution in the absence of a gaseous CO$_2$ feed, and the first example of producing a reduced carbon product other than formate from HCO$_3^-$. We note that the difference in F.E.$_{CO}$ for the N$_2$-saturated 3.0-M KHCO$_3$ solution with the analogous CO$_2$-saturated solution is <2% at every current density between 25 and 100 mA cm$^{-2}$. Electrochemical reduction of the N$_2$-saturated 3.0-M K$_2$CO$_3$ solution was also performed following similar procedures and showed a F.E.$_{CO}$ of 4.0% at 20 mA cm$^{-2}$, and 1.0% at 100 mA cm$^{-2}$ (Figure S2). The relatively low F.E.$_{CO}$ for the CO$_3^{2-}$ solution confirms that CO$_3^{2-}$ is not electrochemically active.

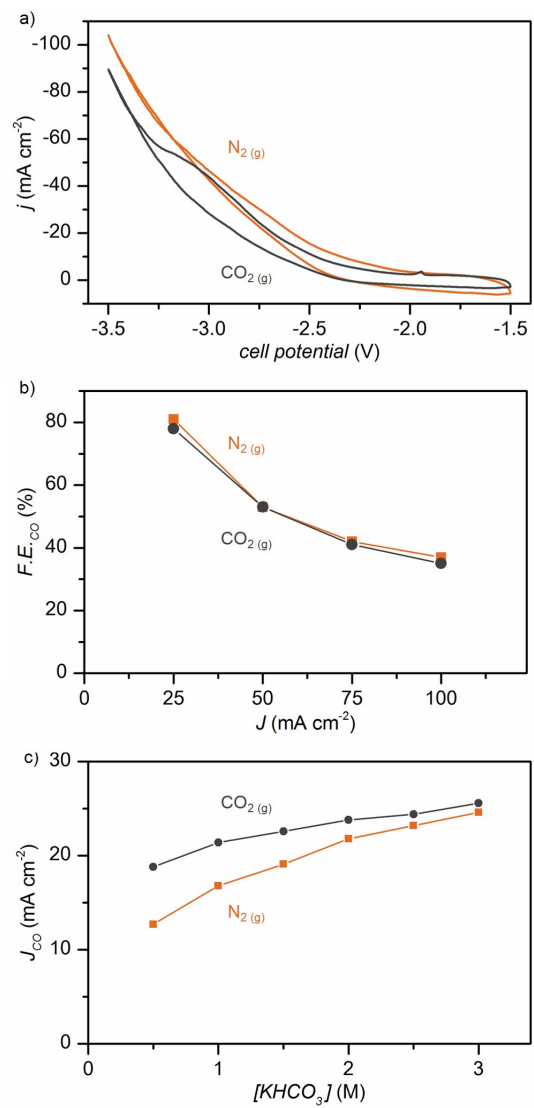

**Figure 2.** Electrochemical reduction of $KHCO_3$ solutions sparged with $CO_{2(g)}$ (black) and $N_{2(g)}$ (orange) in an electrolyzer flow cell containing a BPM show nearly identical behavior: (a) CVs recorded over the -1.5 and -3.5 V range at a scan rate of 100 mV s$^{-1}$ in a 3.0-M $KHCO_3$ solution; (b) F.E.$_{CO}$ measured at constant current densities between 25 and 100 mA cm$^{-2}$ in a 3.0-M $KHCO_3$ solution; (c) $J_{CO}$ as a function of [$KHCO_3$]. Cathode: silver deposited on porous carbon support; anode: nickel foam; catholyte: $KHCO_3$; anolyte: 1.0-M KOH. $J_{CO}$ values were measured at a constant cell potential at 3.0 V in a series of $KHCO_3$ solutions prepared with different $HCO_3^-$ concentrations saturated with $CO_2$ or $N_2$.

The dependence of F.E.$_{CO}$ on [HCO$_3^-$] was investigated by measuring the partial current densities for CO (J$_{CO}$) at a constant cell potential of 3.0 V for a series of N$_2$-saturated and CO$_2$-saturated KHCO$_3$ solutions prepared with HCO$_3^-$ concentrations ranging from 0.5 M to 3.0 M (Figure 2c). The results show that increasing [KHCO$_3$] increases the J$_{CO}$ from 13 mA cm$^{-2}$ at 0.5 M to 25 mA cm$^{-2}$ at 3.0 M. The same solutions bubbled with CO$_2$ show a similar (but less pronounced) increase in J$_{CO}$ from 19 mA cm$^{-2}$ at 0.5-M KHCO$_3$ to 26 mA cm$^{-2}$ at 3.0-M KHCO$_3$. The F.E.$_{CO}$ and total current densities for each electrolyte are provided in Figure S3. The J$_{CO}$ for solutions with <2 M of KHCO$_3$ (where most CO$_2$RR studies are performed) were significantly greater in each case for the CO$_2$-fed electrolytes. The N$_2$-saturated solution yields J$_{CO}$ values comparable with those measured for the CO$_2$-saturated solution at [KHCO$_3$] > 2.5 M. An increase in J$_{CO}$ values with increasing [HCO$_3^-$] (for both the N$_2$-saturated and the CO$_2$-saturated solutions) is consistent with HCO$_3^-$ enhancing the rate of the electrochemical reduction of CO$_2$ through a rapid exchange with CO$_2$[25,26] or suppression of HER with increasing electrolyte pH.[14]

### *Defining the reactions in the flow cell*

Resolving the electrochemically active species in the flow reactor is made challenging by the closed reaction vessel and the dynamic acid-base equilibria that defines the relative concentrations of CO$_2$, HCO$_3^-$, and CO$_3^{2-}$. We therefore designed a set of experiments to confirm that the amount of CO produced from a HCO$_3^-$ stream is governed by three key processes: (i) an acid/base equilibrium of HCO$_3^-$ with protons supplied by the BPM to form CO$_2$ and H$_2$O at the membrane-solution interface; (ii) the reduction of this *in-situ* generated CO$_2$ to CO and OH$^-$; and (iii) the *in-situ* generated OH$^-$ increasing the bulk pH to favor CO$_3^{2-}$ formation, which inhibits CO production (Figure 3a). We operated a flow cell with either a BPM or an AEM at 100 mA cm$^{-2}$ for 2 hours and measured F.E.$_{CO}$ while tracking the concentration of CO$_{2(g)}$ leaving the flow cell ([CO$_2$]$_{outlet}$) and the pH of the bulk catholyte solutions (Figure 3a). We also operated the flow cell containing the BPM, but while circulating KHCO$_3$ without an applied potential (denoted BPM/no electrolysis). These experiments showed that the [CO$_2$]$_{outlet}$ decreased over time with both the BPM and AEM, and that the $[CO_2]_{outlet}$ measured with the BPM system is higher than that measured for the AEM and BPM/no electrolysis (Figure 4a). The formation of $CO_2$ on the cathodic side of the BPM membrane was confirmed by an independent experiment with an H-cell, where operation at 20 mA cm$^{-2}$ produced enough $CO_2$ to exceed the solubility limit of $CO_2$ (Figure 4d). We also observed a correlation of $F.E._{CO}$ to $[CO_2]_{outlet}$ for both the BPM and AEM (Figure 4c), where $F.E._{CO}$ decreases with $[CO_2]_{outlet}$ as a function of time. Finally, the pH of the catholyte increased as a function of time for all three experiments (Figure 4b), where electrolysis accelerated the change in alkalinity.

These collective experimental results support the cathodic reaction chemistry listed in Figure 3b. The higher $[CO_2]_{outlet}$ measured with the BPM than with the AEM is consistent with the BPM supplying a flux of $H^+$ that is available for reaction with $HCO_3^-$ to form $CO_2$ (Figure 3b). This conclusion is further supported by the higher $[CO_2]_{outlet}$ measured when the flow cell with the BPM is subjected to electrolysis (and thus creating a higher $[H^+]$ at the cathode) relative to the experiment without an applied potential. We interpret the linear relationship that exists between $F.E._{CO}$ and the $[CO_2]_{outlet}$ for both membranes as an indication that $CO_2$ is the electrochemically active species; i.e., a higher amount of $CO_2$ available enables more CO to be generated through $CO_2RR$ at the catalyst. This correlation is also consistent with the BPM yielding a higher local $CO_2$ concentration at the catalyst.

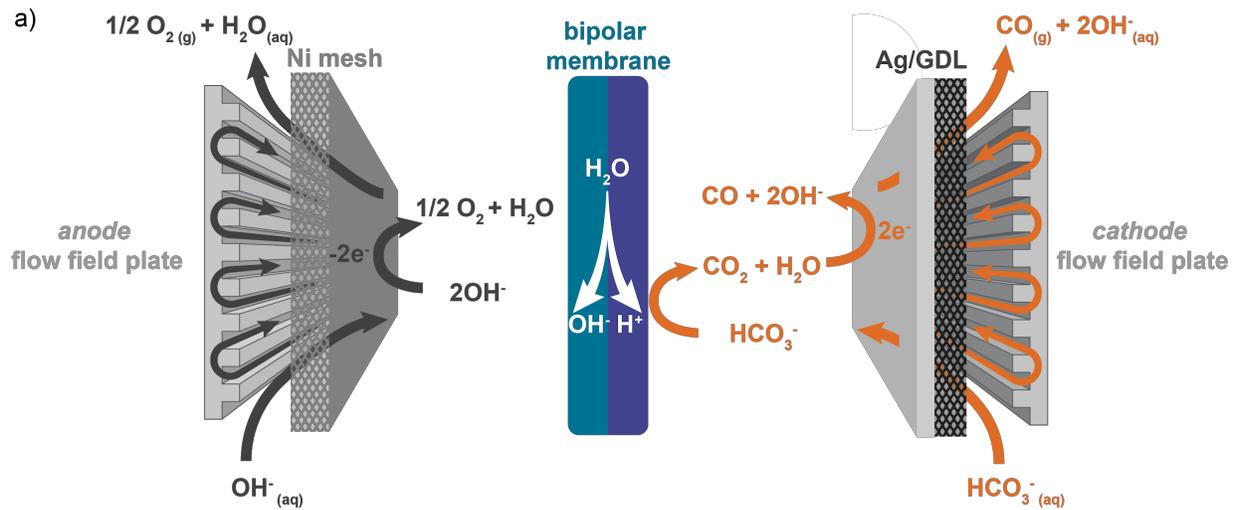

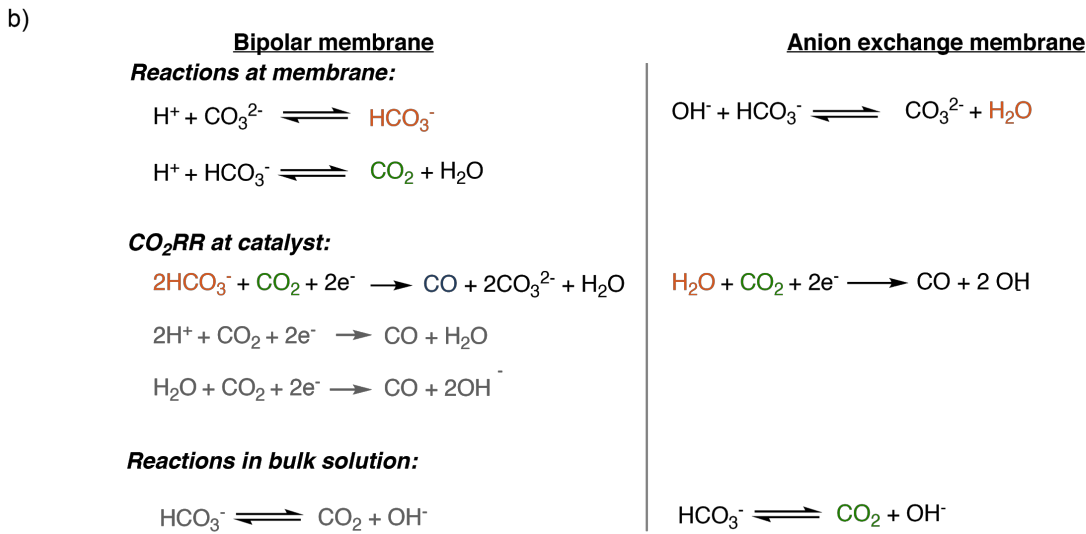

**Figure 3.** a) Schematic representation of the dominant chemical and electrochemical reactions occurring in the flow cell containing a BPM: Reaction of $H^+$ with $HCO_3^-$ to form $CO_2$ at the membrane interface, and the electrochemical reduction of $CO_2$ at the catalyst to form CO and $OH^-$. b) Summary of the reaction chemistry at the membrane and catalyst surfaces and in the bulk solutions in the cathode reaction compartment when a BPM and AEM are used. The most reactive $H^+$ donors (orange) and electrochemically active species (green) that dominate reactivity are highlighted. The reactions indicated in light grey are not expected to govern the reaction chemistry.

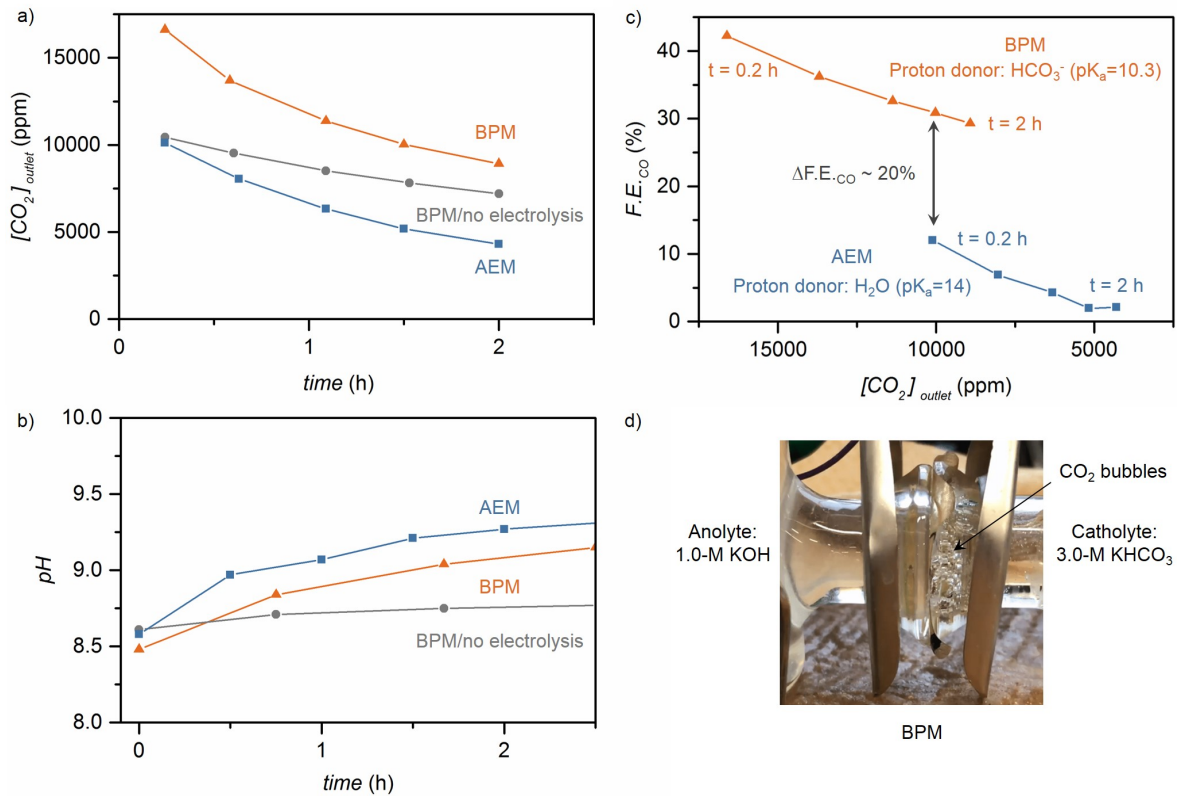

**Figure 4.** Temporal change in (a) $[CO_2]_{outlet}$ and (b) pH during electrolysis of 3.0 M $KHCO_3$ at 100 mA cm$^{-2}$ with a BPM (orange) or AEM (blue). A control experiment, BPM/no electrolysis (grey), was recorded in a flow cell containing a BPM with the circulation of catholyte without an applied potential. (c) $F.E._{CO}$ as a function of $[CO_2]_{outlet}$ during the 2-h electrolysis of a 3.0-M $KHCO_3$ solution at 100 mA cm$^{-2}$ with a BPM (orange) and an AEM (blue). The ~20% difference in $F.E._{CO}$ values at the same $[CO_2]_{outlet}$ is attributed to the difference in pKa values of the indicated H$^+$ donors in the BPM and AEM flow cells. The headspace was purged with a 160 mL/min $N_2$ stream for all experiments. (d) Image of an H-cell where a BPM separates a silver-coated carbon gas diffusion electrode in the cathodic compartment and a Pt mesh anode in the anodic compartment. The formation of $CO_2$ bubbles during the electrolysis of $HCO_3^-$ at a current density of 20 mA cm$^{-2}$ confirm enough $CO_2$ is produced to exceed the solubility limits of $CO_2$ in aqueous media.

The ~20-% higher $F.E._{CO}$ values obtained with the BPM relative to the AEM at the same $[CO_2]_{outlet}$ (e.g., 10,000 ppm) also point to differences in H$^+$ donors available at the $CO_2$RR electrocatalyst surfaces.

Considering that the H$^+$ donors available with the BPM are HCO$_3^-$, H$^+$, and H$_2$O (Figure 3b), we can assume HCO$_3^-$ to be the most active H$^+$ donor because the pKa of 10.3 is lower than that of H$_2$O (pK$_a$ = 14.0), and the high pH would diminish the role of H$^+$. The protons delivered by the BPM would also facilitate the conversion of CO$_3^{2-}$ into HCO$_3^-$, which could then be converted to CO$_2$ through acid/base equilibria. In the case where the AEM is used, the build up of OH$^-$ would deplete HCO$_3^-$ near the membrane surface leaving H$_2$O as the sole H$^+$ donor for CO$_2$RR (Figure 3b). On this basis, we conclude that a higher concentration of H$^+$ donors provided by the BPM than the AEM is responsible for the 20-% difference in F.E.$_{CO}$ at a constant [CO$_2$]$_{outlet}$.

Finally, we are able to rationalize the observed increase in pH of the KHCO$_3$ catholyte solution during electrolysis with both the BPM and AEM systems to be a consequence of OH$^-$ being generated from both HER and CO$_2$RR at the catalyst (Figure 4c). In accordance with previous models,[27,28] this increase in pH would decrease F.E.$_{CO}$ over the course of our electrolysis experiments where the catholyte solutions are not recycled. The higher F.E.$_{CO}$ values could indeed be recovered by replenishing the catholyte with KHCO$_3$ (Figure S4). An additional advantage over the AEM is that the BPM suppresses the increase of the pH of the catholyte during electrolysis (thereby slowing the shift in the equilibrium of carbon species in solution towards catalytically inactive CO$_3^{2-}$; Figures S5-S6) by delivering protons to react with CO$_3^{2-}$ to form HCO$_3^-$.

This collection of experiments point to the conversion of HCO$_3^-$ to CO being enabled by: (i) an acidic region at the membrane interface; and (ii) a basic region at the catalyst. The acidic membrane interface enables CO$_3^{2-}$ to be converted into HCO$_3^-$, and HCO$_3^-$ to be converted into catalytically active CO$_{2(g)}$. The basic pH region at the catalyst layer arises from CO$_2$RR chemistry that generates OH$^-$ concomitant with CO production. An important outcome of this study is the demonstration that the BPM can deliver a local concentration of CO$_2$ to the catalyst that exceeds the solubility limits of CO$_2$ in aqueous media. The ability for the membrane to enable a high concentration of CO$_2$ at the basic catalyst layer offers a strategy for realizing higher CO$_2$RR current densities from aqueous feedstocks. This HCO$_3^-$ reduction system also offers the additional advantage of not requiring a CO$_2$ feed that would acidify the reaction medium and enable HER.

**Conclusions**

We demonstrate that $HCO_3^-$ can be reduced to CO in a flow cell containing a BPM without a supply of $CO_2$ gas to the electrolyte. The observation that a 3.0-M $KHCO_3$ system without a $CO_2$ feed yields a comparable faradaic efficiency for CO with a similar system with a $CO_2$-fed solution was not expected given that it is widely assumed that a $CO_2$ feed is necessary for CO production. The BPM plays a critical role in this reaction chemistry by delivering a $H^+$ flux to the cathode that converts $HCO_3^-$ (and $CO_3^{2-}$) into catalytically active $CO_2$. An important feature of this flow reactor is that the *in-situ* generation of $CO_2$ provides a higher concentration of $CO_2$ at the catalyst surface than the solubility limits of $CO_2$ in aqueous media. This finding offers new opportunities for realizing high current densities with a liquid feed. This work also demonstrates that carbon products other than formate can be generated from $HCO_3^-$ solutions, thereby presenting new opportunities for carbon capture and utilization schemes.

**Experimental Procedures**

**Materials:**

$KHCO_3$ (99%) and $K_2CO_3$ (99%) were purchased from Alfa Aesar. Ag nanopowder (trace metal basis, 99%) and Nafion 117 solution (5 wt%, in a mixture of lower aliphatic alcohols and water) were purchased from Sigma Aldrich. Carbon cloth was purchased from the Fuel Cell Store and cut into desired dimensions with a blade. Nickel foam gas diffusion electrode material was purchased from MTI. BPM (Fumasep FBM) were purchased from FuMA-tech and stored in 1M NaCl Solution.

**Electrode preparation:**

The cathode catalyst ink was prepared by mixing 52 mg of silver nanopowder, 500 µL of DI water, 500 µL of isopropyl alcohol and 70 µl Nafion 117 solution. The catalyst ink was then spray-coated on a 4-cm$^2$ area of carbon cloth and dried under a gentle air stream. Kapton tape (McMaster-Carr) was used as a mask during the deposition process to avoid catalyst being deposited outside the active area of the carbon cloth. The catalyst loading was determined to be 100-120 counts per second by X-Ray fluorescence analysis.

**Electrochemical measurement and product analysis:**

A CH instruments 660D with a picoamp booster was used for all experiments. Electrochemical measurements were made with a two-electrode system with Ni foam as the anode and Ag spray-coated on carbon cloth as the cathode. The anode electrolyte was 1000 mL of 1 M KOH solution delivered by a peristaltic pump at 50 mL/min. The cathode electrolyte was 125 mL of 3 M $K_2CO_3$ or 0.5 ~ 3 M $KHCO_3$ with 0.02 M ethylenediaminetetraacetic acid (EDTA, 99%, Sigma Aldrich) added to remove impurities.[29] The head space of the catholyte solution was purged with $N_2$ (Praxair, 99.9%) or $CO_2$ (Praxair, 99.9%) gas at 160 sccm in a sealed flask with an outlet into the flow cell. The catholyte solution was delivered by another peristaltic pump at 50 mL/min into the flow cell electrolyzer, which was then vented back into the flask. Fresh electrolyte was used for each set of experiment. Samples of the gas headspace in the flask were delivered into a gas chromatograph (GC, Perkin Elmer). The GC was equipped with a packed MolSieve 5A column and a packed HayeSepD column. Argon (Praxair, 99.999%) was used as the carrier gas. A flame ionization detector (FID) equipped with a methanizer was used to quantify CO and $CO_2$ concentrations and a thermal conductivity detector (TCD) was used to quantify $H_2$ concentrations. Control experiments with a two-compartment H-cell were performed wherein the anode (platinum mesh) and cathode (Ag-coated carbon cloth) compartments contained 30 mL of 1 M KOH and 30 mL of 3.0-M $KHCO_3$, respectively, separated by a BPM.

**Acknowledgements:** Funding from The University of British Columbia 4YF Program, Canada Foundation for Innovation, Canada Research Chairs, Quantum Materials and Future Technologies Program, NSERC, and CIFAR is gratefully acknowledged. Jacky Chau fabricated the gas diffusion electrodes used in this study.

**Author contributions**: C.P.B. supervised the project. C.P.B., T.L, M.G., E.W.L, D.A.S. designed the experiments. T.L. performed the proof-of-concept study. D.A.S. designed flow cell assembly and experimental setup. E.W.L. and M.G. defined the reactions in the flow cell. All authors contributed to manuscript writing.

**Declaration of Interests:** The authors declare no competing financial interests.